\documentclass{article}

\PassOptionsToPackage{numbers, compress}{natbib}

\usepackage[final]{neurips_2022}

\usepackage[utf8]{inputenc} 
\usepackage[T1]{fontenc}    
\usepackage{hyperref}       
\usepackage{url}            
\usepackage{booktabs}       
\usepackage{amsfonts}       
\usepackage{nicefrac}       
\usepackage{microtype}      
\usepackage{xcolor}         
\usepackage{graphicx}       
\usepackage{float}
\usepackage{todonotes}
\usepackage{amsmath}
\usepackage{caption}
\usepackage{subcaption}

\title{Looking for Stable Celestial Systems Using Bayesian Optimisation}

\author{%
  Eirik Fladmark\\
  \text{University of Cambridge} \\
  \texttt{ef454@cam.ac.uk} \\
  \And
  Laura B. Justesen\\
  \text{University of Cambridge} \\
  \texttt{lbj25@cam.ac.uk} \\
  \AND
  Teodora Reu\\
  \text{University of Cambridge} \\
  \texttt{tr500@cam.ac.uk} \\
}

\begin{document}

\maketitle

\begin{abstract}

This paper presents a study of the use of numerical simulation and Bayesian optimisation techniques to investigate the dynamics of celestial systems. Initially, the study focuses on Lagrange points in restricted three-body systems where a 2D three-body system simulator is employed to locate the five Lagrange points. An appropriate loss function is developed to capture the gravitational stability of the system, and the stability properties of the different Lagrange points are explored. Additionally, the study investigates how varying the number of variables for the satellite impacts the search for the Lagrange points. Finally, the scope of the study is expanded to explore stable configurations in multi-star systems represented by regular convex \textit{n}-gons. In this case, Bayesian optimisation is used to find suitable settings for the \textit{n}-gon's radius and the stars' velocity vectors, such that the overall system is stable.
\end{abstract}

\section{Introduction}

The \textit{n}-body problem in physics is the study of predicting the motion of \textit{n} celestial objects under the influence of their mutual gravitational attraction \cite{wikipedia_2023}. As the number of objects increases, the mathematical equations describing their motion become increasingly complex, making the problem challenging to solve. Methods such as numerical integration \cite{merritt1985numerical} and computer simulations \cite{aarseth2003gravitational} have been used to approximate solutions to the \textit{n}-body problem. Advances in high-performance computing have improved the accuracy of simulations of large \textit{n}-body systems, such as galaxies and galaxy clusters. Despite this, the problem remains a challenging area of research.

In this study, we investigate the \textit{n}-body problem by using simulators with Bayesian optimisation to find stable configurations. Section \ref{section:lagrange} focuses on locating Lagrange points, while Section \ref{section:n-gon} presents a search for stable configurations in multi-star systems. Finally, we conclude our project in Section \ref{section:conclusion}.




\section{Locating Lagrange Points} \label{section:lagrange}
First, we wanted to simulate a well-known phenomenon in restricted three-body systems: Lagrange points. The Lagrange points are equilibrium points for objects of negligible mass when affected by two orbiting bodies of much greater mass \cite{wiki:lagrange}. At these points, the gravitational forces balance the centrifugal force \cite{weisstein}. There are five different Lagrange points (L1--L5) as shown in Figure \ref{fig:lagrange}. L1, L2, and L3 are located on the straight line going through the centres of the two massive bodies, while L4 and L5 are respectively located as the third vertex of an equilateral triangle between that point and the centres of the two massive bodies \cite{wiki:lagrange}.

\subsection{Stability of Lagrange Points}
\begin{figure}[ht]
\includegraphics[scale=0.7]{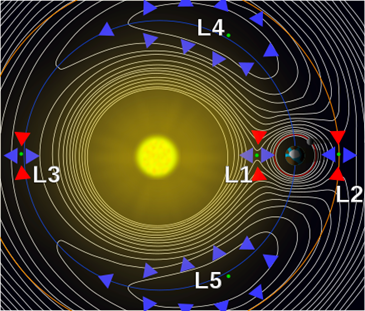}
\centering
\caption{Locations of the five Lagrange points and their stability regions. Source: \cite{wiki:lagrange}}
\centering
\label{fig:lagrange}
\end{figure}

Despite being equilibria, L1 and L2 are saddle points and dynamically unstable. If a satellite located at one of these points deviates from the equilibrium, this deviation will grow exponentially with time. Thus, the satellite will leave the equilibrium unless force is exerted to correct its course \cite{cornish1998}. Similarly, L3 is a weak saddle point, making its orbit exponentially unstable. However, its \textit{e}-folding time (the time it takes for an exponentially growing quantity to increase by a factor of \textit{e} \cite{wiki:efolding}) is longer, making it slightly more stable \cite{cornish1998}. Lastly, L4 and L5 are the only stable equilibria. The stability of these points is caused by the Coriolis force. If a satellite is located near L4 or L5, it will have a tendency to move away. However, this will result in an increase in its speed, which in turn causes the Coriolis force to lead the satellite into an orbit around the equilibrium point \cite{cornish1998}. The resulting orbit is slightly bean-shaped (as seen in Figure \ref{fig:lagrange}) \cite{wiki:lagrange}. It should be noted that this stability requires the ratio between the mass of the largest body and the mass of the second-largest body to be greater than 24.96 \cite{wiki:lagrange}.

\subsection{Methodology}

We wanted to locate the five Lagrange points using a 2D three-body system simulator and Bayesian optimisation. For the simulation, we use the Sun and the Earth as the two large bodies in the system, while the third (and smallest) body is a small satellite. In particular, our system replicates the relationship between the Sun and the Earth in terms of relative weights and distance (the Satellite is of negligible mass). We will therefore be referring to the different bodies in the system as the Sun, the Earth, and the Satellite. The Sun is located at (0, 0). However, there was no readily available loss function, which captures gravitational stability. Therefore, the search for the Lagrange points also included a search for an appropriate loss function. Additionally, we also experimented with what knowledge we could incorporate in the search space of the Bayesian optimisation. Lastly, simulations were run for between 1,000 and 3,000 iterations (depending on the specific experiment), which is equivalent to between around 5.5 and 17 orbits around the Sun.

Several experiments were performed in which we varied the number of variables (for the Satellite) between 2 and 4 (where 4 would be \textit{x}-position, \textit{y}-position, \textit{x}-velocity, and \textit{y}-velocity). When using our approximated loss function, using more than 2 variables resulted in poor performance for the search. Thus, we had to compensate with knowledge about the system. We, therefore, utilised the following two things we knew about the system:

\begin{enumerate}
  \item The approximate geometric positions of the Lagrange points are known. L1, L2, and L3 lie on the straight line going through the centres of the Sun and the Earth (as mentioned earlier), while L4 and L5 are located on the Earth’s orbit around the Sun. 
  \item The positions of the Lagrange points are constant relative to the Sun and the Earth, and the orbits are almost circular (slightly oval). Therefore, the \textit{x}-velocity and \textit{y}-velocity can be calculated using one variable (the speed) and cosine (for \textit{y}-velocity) or sine (for \textit{x}-velocity).
\end{enumerate}

\subsection{Locating L3 and Formulating a Loss Function}

Since L3 is the most isolated of the Lagrange points (see Figure \ref{fig:lagrange}), this was the first point we tried to locate. However, before we could do this, we first needed to formulate a suitable loss function that could represent the stability of a point in space. Our first idea was based on the distance between the Sun and the Satellite. We know that L3 is an equilibrium, and whilst it might not be perfectly stable (see earlier discussion of stability), it should be more stable than the points surrounding it. As mentioned previously, we make the greedy approximation that the Earth and L3 move in circular orbits around the Sun. Thus, the loss was calculated as the biggest deviation from the Satellite's initial distance from the Sun during the 1,000 iterations of the simulation. However, since the Earth's orbit is not perfectly circular, we will get some loss in all instances where the Earth is not exactly 200 units from the Sun, which is suboptimal. When using this loss to find L3, we observed that it selected other points that appeared to follow slightly more circular orbits. However, these satellites would either slow down or speed up and thus not stay opposite the Earth (with regards to the Sun) as objects in L3 do.

This observation led us to another realisation: the relative angles of the system are constant. The Earth and L3 will therefore always have a 180° angle between them (with the Sun as the angle's vertex). Thus, adding the biggest deviation from the initial Satellite-Sun-Earth angle to the loss should make it easier to avoid selecting satellites that orbit too slowly or too fast. After some experimentation, we found that this combined loss (i.e. distance and angle loss) yielded much better results. Furthermore, since L1--L3 are very sharp minima, we restricted the search space to be around the area where it should be when performing Bayesian optimisation. In the case of L3, we know the distance from the Sun to the Earth (200 in the simulation), so we searched ±10\% of that distance (on the horizontal axis), on the opposite side of the Sun.

\begin{figure}[t]
\includegraphics[width=\textwidth]{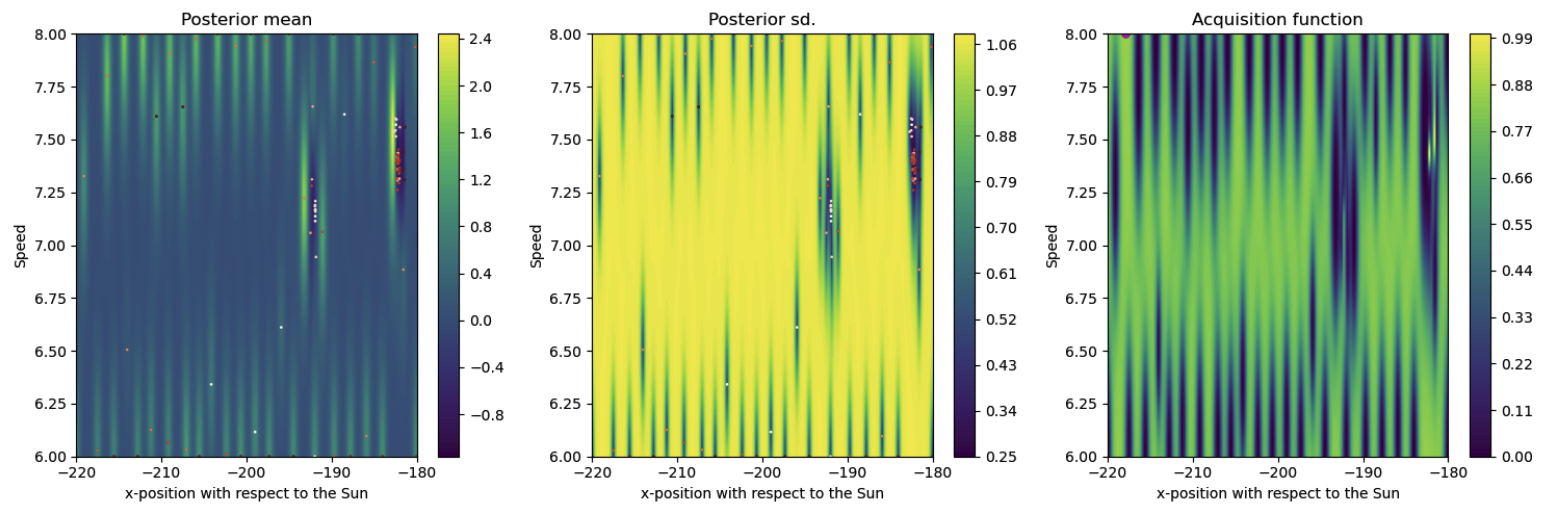}
\centering
\caption{Results for L3 using distance-based loss function.}
\centering
\label{fig:l3-bad-loss}
\end{figure}

\begin{figure}[t]
\includegraphics[width=\textwidth]{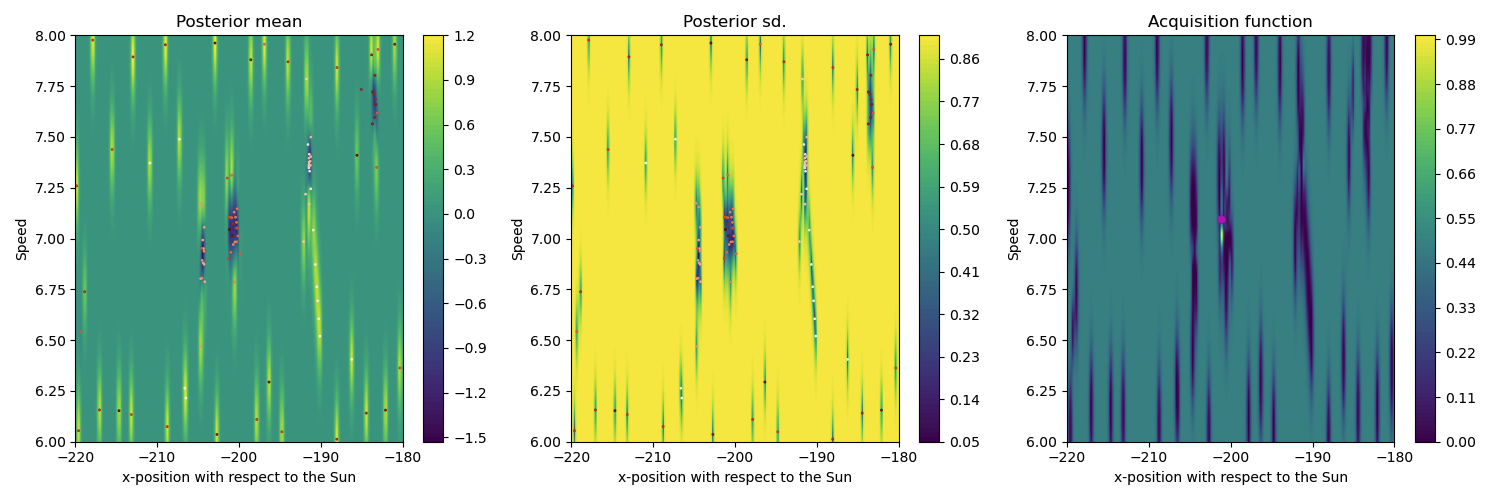}
\centering
\caption{Results for L3 using combined loss function. The best point is located at \(-\)199.33 with a speed of 7.10.
}
\centering
\label{fig:l3-good-loss}
\end{figure}

Figure \ref{fig:l3-bad-loss} shows the results for the distance-based loss function, while Figure \ref{fig:l3-good-loss} shows the results for the combined loss function. We observe that the distance-based loss function is ill-suited, as we do not find L3. On the other hand, the combined loss function performs well and seems to find L3 exactly opposite the Earth (\(-200\)) with a speed of around \(-7\) (minus indicates direction), as seen in the leftmost plot. However, this result should be taken with a grain of salt. While it is true that L3 is located in this area, it is also likely that this position and speed are suitable for general orbits. For example, the Earth orbits with a corresponding position and speed on the other side of the Sun.

\subsection{Locating L1 and L2}

We then applied the combined loss function to L1 and L2. We know that the points are located on the straight line going through the centres of the Sun and the Earth (see Figure \ref{fig:lagrange}) and that they are both around 0.01 AU from the Earth \cite{nasa:webb-orbit}. Due to their proximity to the Earth, they are more prone to be affected by the gravity of the Earth (or the Sun).

The results of running Bayesian optimisation with a similar search space as for L3 (just on the opposite side of the Sun) were unstable (when repeating the experiment), and we were unable to locate L1 and L2. Further investigation showed that the areas selected by Bayesian optimisation, though different from L1/L2, resulted in somewhat stable satellite orbits. Furthermore, these points were not as sharp as L1/L2. L1 and L2 are sharp minima whose surrounding points are very unstable (resulting in the satellite being shot out of orbit or consumed by the Sun). Bayesian optimisation is therefore more likely to find a bigger and (somewhat) stable area.

\begin{figure}[t]
\includegraphics[width=\textwidth]{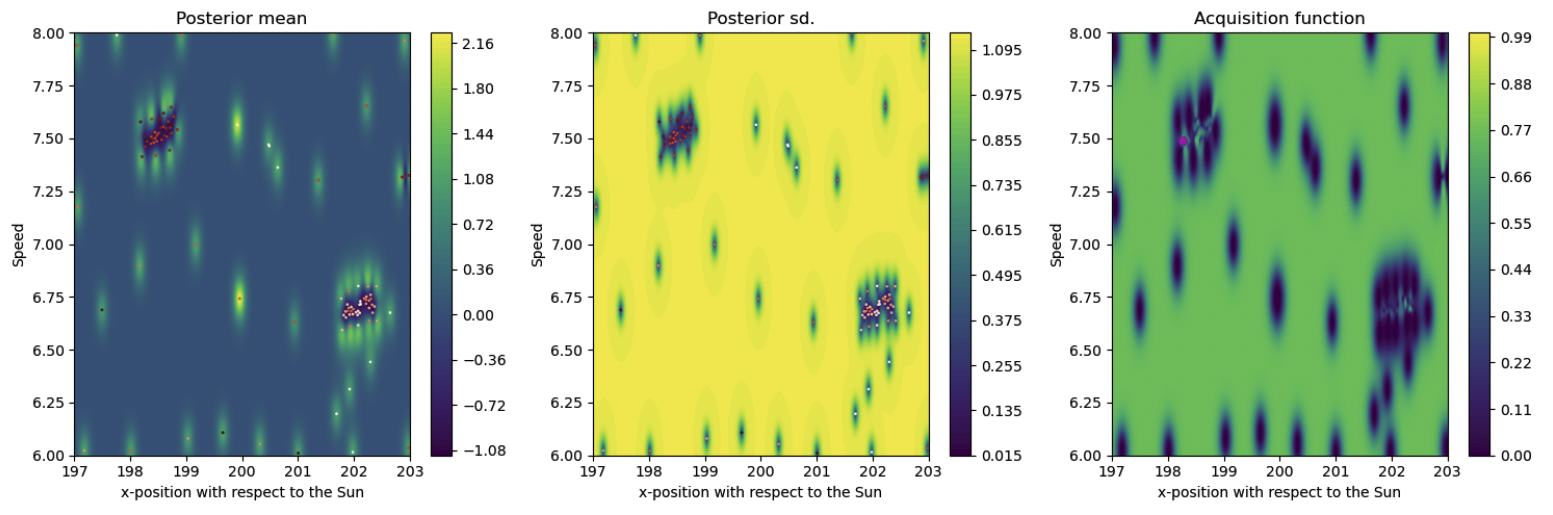}
\centering
\caption{Results for L1 and L2 on a restricted search space. The best point was at 198.74 units distance with a speed of 7.52.}
\centering
\label{fig:l1-l2-restricted}
\end{figure}

Figure \ref{fig:l1-l2-restricted} shows the results of restricting the search space to exclude these areas. At first glance, we observe two groupings at the correct locations for L1 and L2 (at 198 and 202). However, their respective speeds are opposite of what we would expect. Since the angular velocity is the same for objects in L1 and L2, L2 (which is further away from the Sun) should have a higher speed than L1. Furthermore, the difference in speed should be less than observed, since L1 and L2 are relatively close to each other. Further investigation indicated that the satellite enters an orbit around the Earth, which results in a low angle loss (for the Earth-Sun-Satellite angle).

The loss function was modified in an attempt to address this problem. First, we tried to change the angle's vertex from the Sun to the Earth, such that orbits around the Earth would increase the loss. However, this still resulted in all satellites either orbiting the Earth or being sent out of orbit completely. Second, we recorded the number of times a satellite orbited the Earth and added this to the loss. However, the minimum loss in this case still resulted in a satellite orbiting the Earth. The extreme sensitivity of L1 and L2 illustrates the issues with Bayesian optimisation when working with sharp minima. These results might also put our discovery of L3 in doubt. However, since the surroundings of L3 are less sensitive, it should be easier to locate than L1 and L2. 

\subsection{Locating L4 and L5}

\begin{figure}[t]
\includegraphics[width=\textwidth]{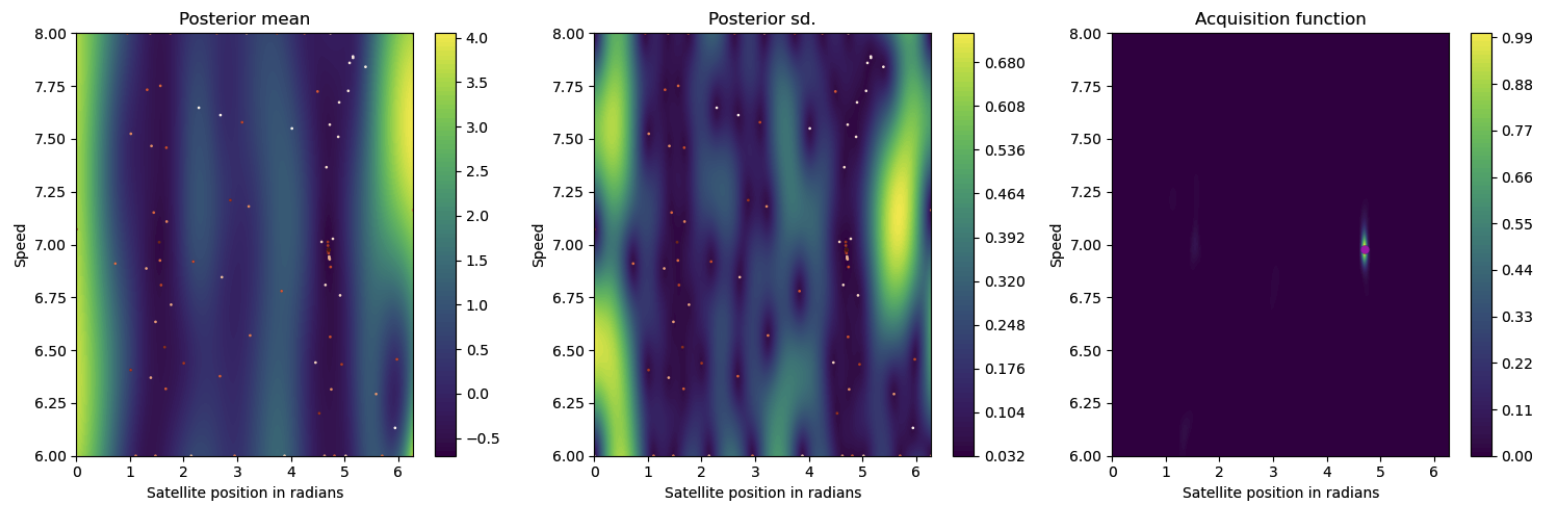}
\centering
\caption{Results for L4 and L5. The best point was at 4.70 radians with a speed of 6.97.}
\centering
\label{fig:l4-l5}
\end{figure}

The stable nature of L4 and L5 should make them easier to locate. As mentioned earlier, we only use two variables when searching: satellite position and speed. Here we search for the satellite position on a circle with a radius of 200 around the Sun (this radius is equal to the Earth's initial distance from the Sun). The position is given in radians, with 0 radians equal to the Earth's initial position. We restrict the search space like this since we know that L4 and L5 should lie approximately on the Earth's orbit. This works since L4 and L5 are stable equilibria. If they were unstable and sharp minima, they would need to lie exactly on the Earth's orbit for this to work. However, due to their stability, we expect fairly good performance, even if we are slightly off the exact location of L4 and L5.

The results are shown in Figure \ref{fig:l4-l5}. As expected, L4 and L5 were the easiest Lagrange points to locate due to their less sharp minima. We observe that the loss valleys occur just after 1 and just before 5 radians, which is roughly where we expect them to be, considering that L4 and L5 should lie 60° above and below the Earth respectively (60° = \(\pi \div 3 \approx 1.05\) radians, while \(-60\)° (or 300°) = \(\pi \div 3 \times 5 \approx 5.24\)). We also note that the velocity does not need to be one exact value as any deviation will still keep the satellite around due to the stable nature of these two points.

\subsection{Conclusion and Future Directions}
The experiments presented in this section show both the utility and limitations of using Bayesian optimisation in search of equilibria and stable points in space. The most time-consuming task was experimenting with the loss functions for L1 and L2. These points are close to the Earth and are both very sharp minima---something Bayesian optimisation has trouble finding.

In the future, it would be interesting to replace the circular assumption by creating a periodic loss function, which starts after the first period and then calculates the shape loss between the second and third periods, the third and fourth periods, and so on. It could also be beneficial to keep track of the position and velocity of the satellite at different points since a satellite in a stable orbit should have the same speed every time it is in one specific location. However, the methods would optimally be used to try and find stable equilibria, as Bayesian Optimisation does not seem like the best tool to find unstable equilibria.

Overall, a demonstration of how loss functions and Bayesian optimisation can be used to map out areas of space has been presented, and we now move on to finding some less established stable environments.

\section{Building Stable \textit{n}-gon Star Systems} \label{section:n-gon}

\begin{figure}[t]
\includegraphics[scale=0.32]{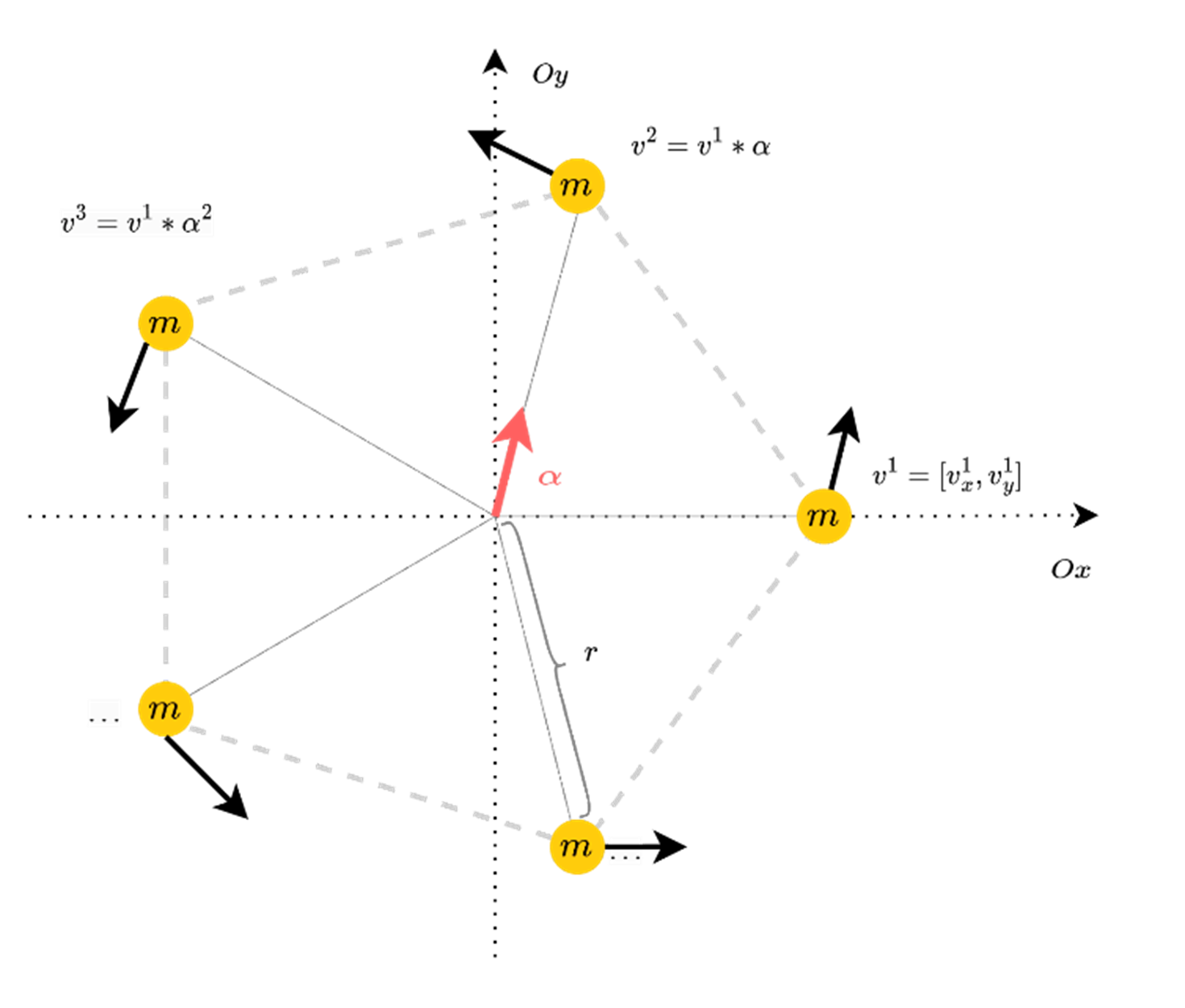}
\centering
\caption{Visual representation of a 5-gon star system.}
\centering
\label{fig:n-gon-diagram}
\end{figure}

After using Bayesian optimisation to locate the Lagrange points, we wanted to investigate how Bayesian optimisation might be used to find other stable systems. We, therefore, decided to search for stable configurations in \textit{n}-gon star systems (Figure \ref{fig:n-gon-diagram} shows a 5-gon star system).

An \textit{n}-gon is an \textit{n}-sided polygon \cite{wiki:polygon}, and the \textit{n}-gon star system is represented by a regular convex \textit{n}-gon, such that all sides are of equal length and all interior angles are equal and less than \(\pi\). The \textit{n}-gon is situated in 2D space with its centre at coordinate \((0,0)\). \textit{n} stars (each with a mass of 10,000) are then placed with their centres on the vertices of the \textit{n}-gon. Every vertex is at a distance (or radius) of \textit{r} from the centre. The first vertex (and star) is fixed on the horizontal axis (see Figure \ref{fig:n-gon-diagram}), while the others are placed according to the restrictions of the \textit{n}-gon (to maintain regularity and convexity).

The goal is then to use Bayesian optimisation to find a setting for the radius of the \textit{n}-gon (\textit{r}) as well as velocity vectors (\(v_x\), \(v_y\)) for each star, such that the overall system is stable. The motivation for finding stable \textit{n}-gon star systems was that this could be a starting point for further interesting investigations. For example, we could then simulate such solar systems with planets orbiting the \textit{n} stars. We wonder whether there might exist "Lagrange points" for such \textit{n}-star solar systems, and whether such \textit{n}-star "Lagrange points" could be a generalisation of the normal Lagrange points.

The search space for this problem consists of \(2n+1\) variables since we need to find velocity vectors (\(v_x\), \(v_y\)) for each star as well as the radius of the overall system. In order to check configurations with a large number of stars, it would be desirable to simplify the search space. It is possible to reduce the search space to three dimensions regardless of the value of \textit{n}. The search space will then consist of the radius (\textit{r}) and the velocity vector of the first star ($v^{1}=[v^{1}_x,v^{1}_y]$). The velocity vectors of the remaining stars ($v^{k}$ for \(k \in \{2,...,n\}\)) can be calculated as $v^{k} = v^{1} * \alpha^{k-1}$, as shown in Figure \ref{fig:n-gon-diagram}. \(\alpha\) is a complex number with length 1 and angle \(2\pi/n\) where \textit{n} is the number of vertices (see Equation \ref{eq:alpha} for the formal definition). It is used to rotate the velocity of the first star appropriately. This simplification of the search space allows us to study settings with large \textit{n}.

\begin{equation} \label{eq:alpha}
    \alpha = \cos(2\pi/n) + i \sin(2\pi/n)
\end{equation}

In the following sections, we will present some of the configurations found by Bayesian optimisation and the loss functions used. The loss function defines the stability of the system. For example, one possible loss function would be how far a star has moved from the centre \((0,0)\). In the search phase, the simulation is run for 10,000 iterations, which is equivalent to 55 Earth years (or Earth cycles around the Sun). For \textit{n}-gons with \textit{n} greater than 3, we also look at the evolution of the system over 100,000 iterations, which is equivalent to 550 Earth years. As we increase the running time of the experiment, we expect the returned solutions to be more stable.

\subsection{2-gon Star Systems}

First, we examined star systems consisting of two stars. The search space was restricted as follows: the radius \textit{r} was varied between 150 and 250, while the velocity vectors for the first star were varied between \(-5\) and \(5\).

We used two different loss functions. Let $s^{t}_1$ and $s^{t}_2$ be the positions in space of the two stars at time $t$, where $t \in \{0,1, ...T\}$ represents the current iteration. The first loss function (\(L_1\)) measures the maximum distance between the two stars as defined in Equation \ref{eq:loss-1} below:

\begin{equation} \label{eq:loss-1}
    L_{1} = \max\limits_{t \in T} dist(s^{t}_1, s^{t}_2)
\end{equation}

where $dist(a,b)$ calculates the Euclidean distance between two points $a, b \in R^n$ (in this case $n=2$). The second loss function (\(L_2\)) looks at the difference between a star's current and initial radius, as defined in Equation \ref{eq:loss-2} below:

\begin{equation} \label{eq:loss-2}
    L_{2} = \max_{\substack{t \in T, \\ i \in \{1,2\}}} ||{\Vec{s^{t}_i}}|-|{\Vec{s^{0}_i}}||
\end{equation}

The intuition behind \(L_2\) is that we are trying to force the stars to lie on an orbit that deviates the least from a circular orbit around the centre with radius $r = |{\Vec{s^{0}_1}}| = |{\Vec{s^{0}_2}}|$.

\begin{figure}
     \centering
     \begin{subfigure}[b]{0.24\textwidth}
         \centering
        \includegraphics[width=\textwidth]{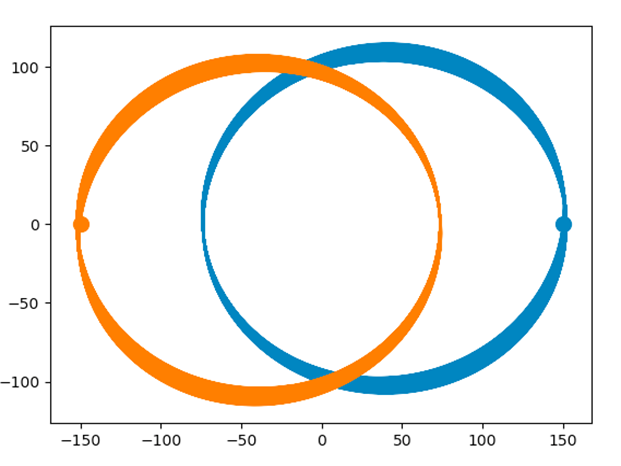}
         \caption{Using \(L_1\)}
         \label{fig:2-gon-l1-1}
     \end{subfigure}
     \hfill
     \begin{subfigure}[b]{0.24\textwidth}
         \centering
         \includegraphics[width=\textwidth]{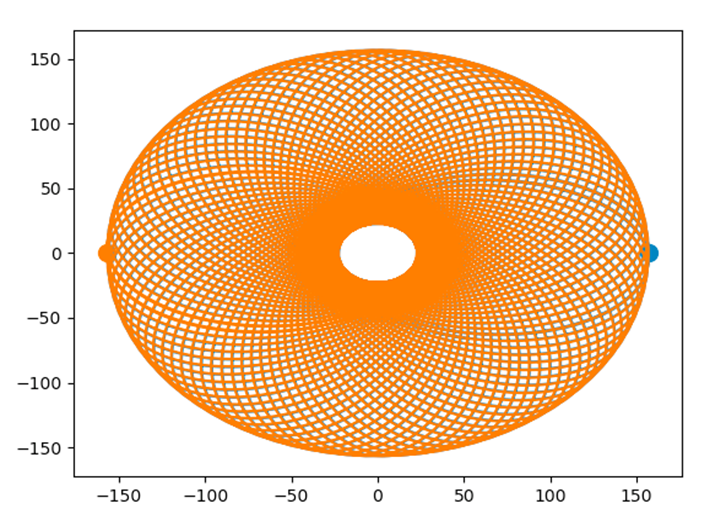}
         \caption{Using \(L_1\)}
         \label{fig:2-gon-l1-2}
     \end{subfigure}
     \hfill
     \begin{subfigure}[b]{0.24\textwidth}
         \centering
         \includegraphics[width=\textwidth]{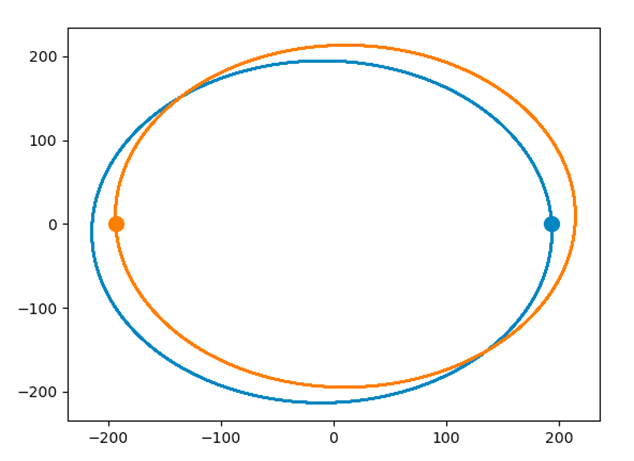}
         \caption{Using \(L_2\)}
         \label{fig:2-gon-l2-1}
     \end{subfigure}
     \hfill
     \begin{subfigure}[b]{0.24\textwidth}
         \centering
         \includegraphics[width=\textwidth]{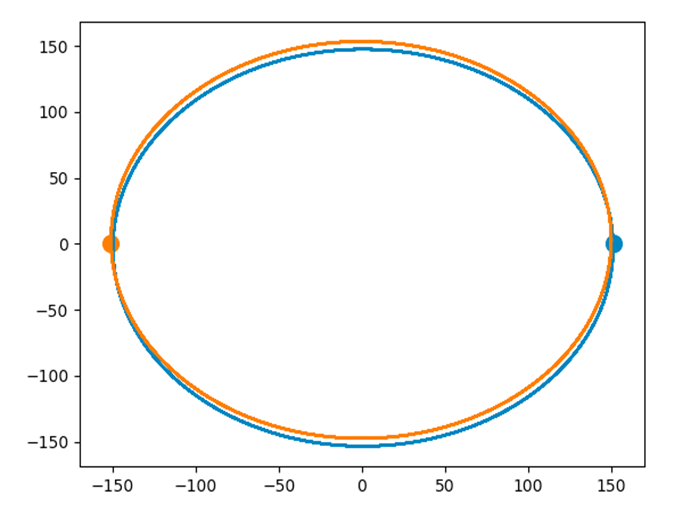}
         \caption{Using \(L_2\)}
         \label{fig:2-gon-l2-2}
     \end{subfigure}
        \caption{Stable configurations found for 2-star systems.}
        \label{fig:2-gon}
\end{figure}

When using the aforementioned configurations, Bayesian optimisation resulted in several stable systems. The orbits of four of these systems are shown in Figure \ref{fig:2-gon}, with Figures \ref{fig:2-gon-l1-1} and \ref{fig:2-gon-l1-2} using \(L_1\) and Figures \ref{fig:2-gon-l2-1} and \ref{fig:2-gon-l2-2} using \(L_2\). The coloured dots indicate the initial positions of the stars. When comparing the orbits for the two loss functions, we notice that \(L_2\) forces the orbits' centres to be very close to the overall centre of the coordinate system. Furthermore, though the orbits look slightly oval in the figure (due to the scaling of the axes), upon further inspection, we observe that the trajectories (for \(L_2\)) are, in fact, circular.

\subsection{3-gon Star System}

For \textit{n}-gons with $n \geq 3$, the previously described loss functions (\(L_1\) and \(L_2\)) did not return adequately stable systems. We, therefore, propose a third loss function 
(\(L_3\)) based on the area of the shape formed by the stars in space at a time \textit{t}. It measures the maximum deviation from the initial area. The formula for this loss is given in Equation \ref{eq:loss-3} below:

\begin{equation} \label{eq:loss-3}
    L_{3} = \max_{\substack{t \in T, \\ i \in \{1,2\}}} |{\mathcal{A}_{s_1^{t},s_2^{t},...s_n^{t}}}-{\mathcal{A}_{s_1^{0},s_2^{0},...s_n^{0}}}|
\end{equation}

where $\mathcal{A}_{a_1,a_2,..., a_n}$ is the surface area of the shape produced by connecting the points $a_1,a_2,..., a_n$ in space. This is calculated using the Shapely Python library.

\begin{figure}[ht]
\includegraphics[scale=0.5]{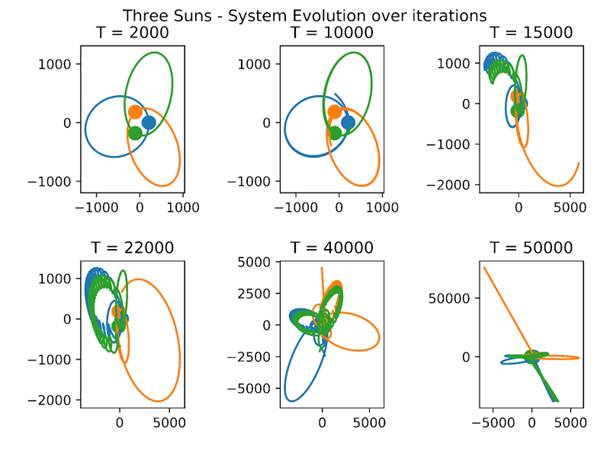}
\centering
\caption{Evolution of a 3-star system over 50,000 iterations.}
\centering
\label{fig:3-gon-evolution}
\end{figure}

Finding stable configurations for three stars was more challenging than expected. A configuration, which appears stable early on, might destabilise later. This problem is illustrated in Figure \ref{fig:3-gon-evolution}. It shows one of the best solutions found by Bayesian optimisation (\(T=\) 2,000). At this point in time, the system looks completely stable. However, after 10,000 iterations, it starts to destabilise. It is interesting to observe how the system exhibits somewhat symmetric properties. For example, (at \(T=\) 15,000 and \(T=\) 22,000) we can see that the orange star moves in an orbit with around double the diameter of the "orbits" of the blue/green stars. It appears as if the energy in the system has been evenly split between the orange star and the green/blue stars. Later on (\(T=\) 40,000), we observe that the event at \(T=\) 22,000 has been mirrored (resulting in a butterfly-like shape). Eventually, the stars leave the canvas completely (\(T=\) 50,000). It should be noted that although the images in Figure \ref{fig:3-gon-evolution} may give the impression of a 3D space, the stars only move in 2D.

In Figure \ref{fig:3-gon-evolution}, the stars are initially located close to each other. This is because the radius domain was restricted to between 200 and 300. We then decided to search for stable configurations in the following different radius intervals: 100--250, 200--250, 250--500, 500--750, 750--1000, 1000--2000, and 1500--2500. To further simplify the search, we reduce the search space for the velocity by initializing \(v^1_x\) to 0.

\begin{figure}
     \centering
     \begin{subfigure}[b]{0.3\textwidth}
         \centering
        \includegraphics[width=\textwidth]{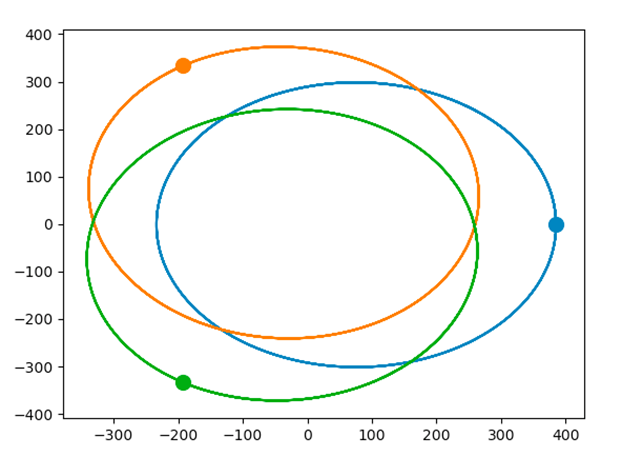}
         \caption{}
         \label{fig:3-gon-1}
     \end{subfigure}
     \hfill
     \begin{subfigure}[b]{0.3\textwidth}
         \centering
         \includegraphics[width=\textwidth]{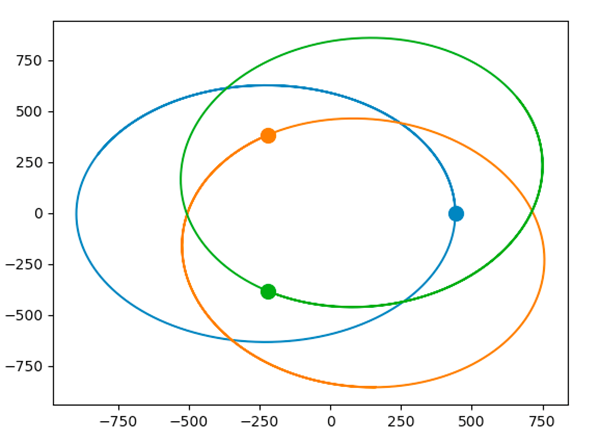}
         \caption{}
         \label{fig:3-gon-2}
     \end{subfigure}
     \hfill
     \begin{subfigure}[b]{0.3\textwidth}
         \centering
         \includegraphics[width=\textwidth]{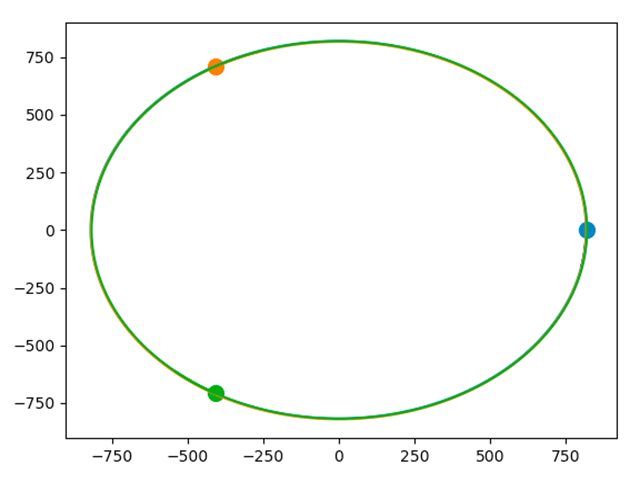}
         \caption{}
         \label{fig:3-gon-3}
     \end{subfigure}
        \caption{Stable configurations found for 3-star systems.}
        \label{fig:3-gon}
\end{figure}

Using these search-space settings, Bayesian optimisation found three types of "stable" configurations. Figure \ref{fig:3-gon} shows the stable configurations at iteration 2,000 where the stars have completed at least two cycles on the found orbits. We identify three distinct configurations here: one where the orbit of any star does not contain the initial position of the other stars (Figure \ref{fig:3-gon-1}), one where each star's orbit contains the initial positions of the other stars (Figure \ref{fig:3-gon-2}), and finally one where all the stars move in the same orbit (Figure \ref{fig:3-gon-3}).

\subsection{4-gon Star System}

\begin{figure}
     \centering
     \begin{subfigure}[b]{0.3\textwidth}
         \centering
        \includegraphics[width=\textwidth]{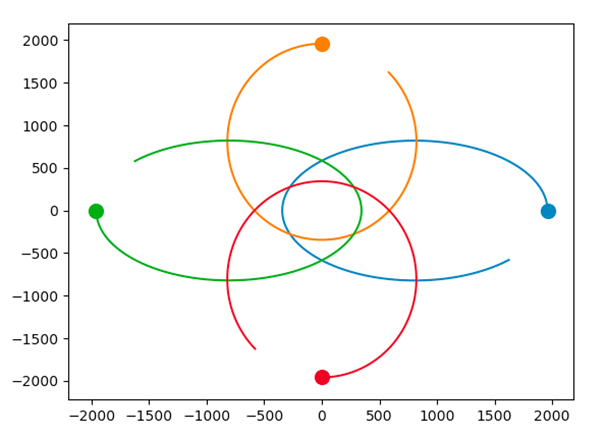}
         \caption{}
         \label{fig:4-gon-1}
     \end{subfigure}
     \hfill
     \begin{subfigure}[b]{0.3\textwidth}
         \centering
         \includegraphics[width=\textwidth]{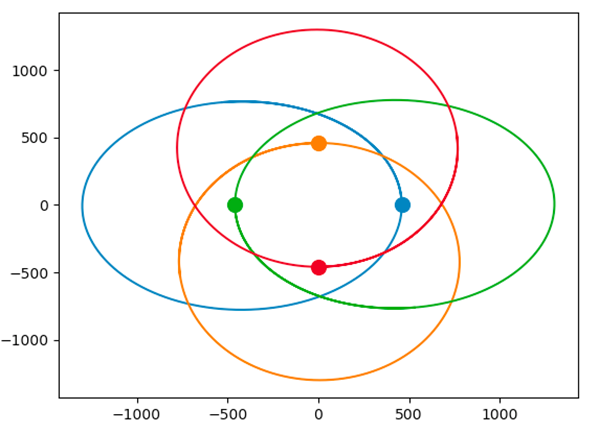}
         \caption{}
         \label{fig:4-gon-2}
     \end{subfigure}
     \hfill
     \begin{subfigure}[b]{0.3\textwidth}
         \centering
         \includegraphics[width=\textwidth]{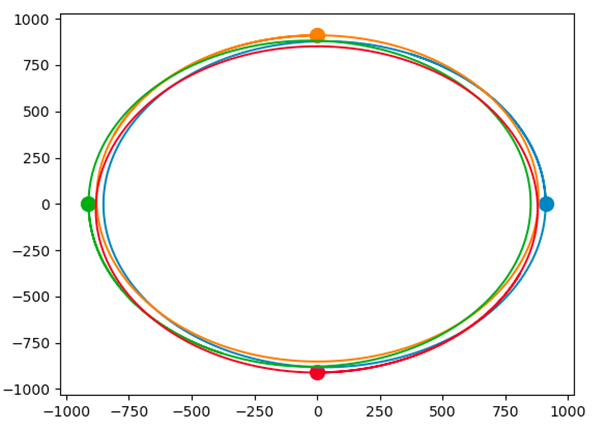}
         \caption{}
         \label{fig:4-gon-3}
     \end{subfigure}
        \caption{Stable configurations found for 4-star systems.}
        \label{fig:4-gon}
\end{figure}

For 4-gon star systems, we used the same settings as when looking for stable 3-gon star systems. We also used the same area-based loss function (\(L_3\)). Figure \ref{fig:4-gon} show three stable configurations found by Bayesian optimisation. We observe that these systems have the same distinct structure as the configurations found for 3-gon star systems (Figure \ref{fig:3-gon}).

\subsection{\textit{n}-gon Star Systems with \textit{n} > 4}

\begin{figure}
     \centering
     \begin{subfigure}[b]{0.29\textwidth}
         \centering
        \includegraphics[width=\textwidth]{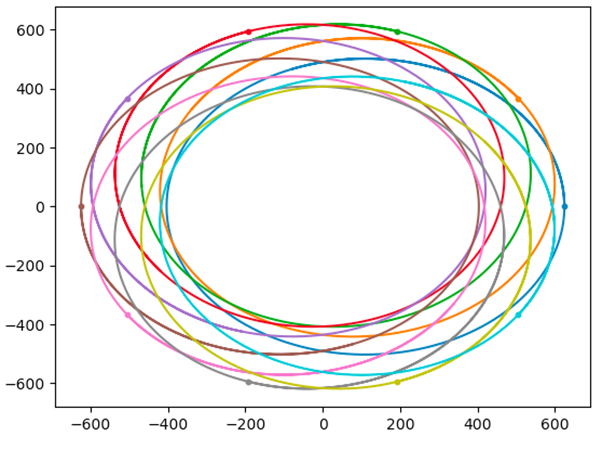}
         \caption{10-star system}
         \label{fig:10-gon}
     \end{subfigure}
     \hfill
     \begin{subfigure}[b]{0.32\textwidth}
         \centering
         \includegraphics[width=\textwidth]{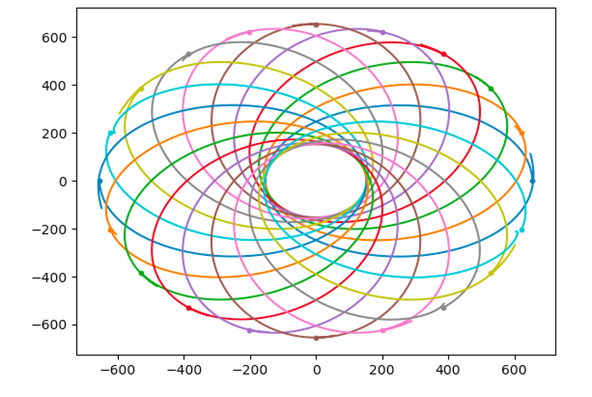}
         \caption{20-star system}
         \label{fig:20-gon-1}
     \end{subfigure}
     \hfill
     \begin{subfigure}[b]{0.3\textwidth}
         \centering
         \includegraphics[width=\textwidth]{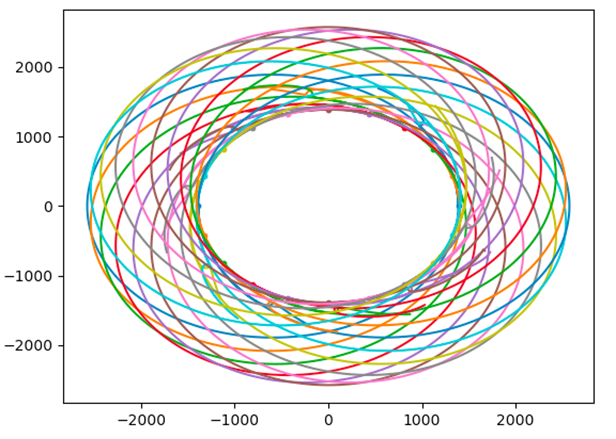}
         \caption{20-star system}
         \label{fig:20-gon-2}
     \end{subfigure}
        \caption{Stable configurations found for \textit{n}-gon star systems with \(n > 4\).}
        \label{fig:n-greater-than-4}
\end{figure}

When more stars are added to the system, Bayesian optimisation finds solutions further away from the centre of the coordinate system, which poses a problem. As we move further away from the centre in the same number of iterations, the stars will complete fewer orbits since their paths increase in length. A future step for this \textit{n}-gon star system problem is to find the mathematical reasoning behind this phenomenon. To end this investigation, we here include plots for other \textit{n}-gon structures (where the mass of the stars has been changed to 200). These plots are shown in Figure \ref{fig:n-greater-than-4}. Figure \ref{fig:10-gon} shows the orbits of a system of 10 stars with the initial positions on the outside. Figure \ref{fig:20-gon-1} shows the orbits of a system of 20 stars where the initial positions are still located on the outside. Finally, Figure \ref{fig:20-gon-2} shows the orbits of a system of 20 stars with the initial positions on the inside.

\subsection{Discussion and Future Directions}

It is hard to find stable star systems and planetary systems. Using Bayesian optimisation to approach this problem can be both exciting and fun. However, Bayesian optimisation can only ensure that it has found a truly stable system if it is able to look to the end of time, which is not possible in simulations. Furthermore, even when using relatively simple loss functions that punish deviations in the system (e.g. \(L_1\), \(L_2\), and \(L_3\)), the loss landscape is very noisy, and the solutions occur at very sharp minima.

In this section, we have seen examples of stable systems with multiple stars. Furthermore, we have gained an intuition of the three configurations that appear to be possible for each \textit{n} (see Figure \ref{fig:3-gon}). Moving forward, we believe it would be advisable to take a more mathematical approach. After finding the mathematical solution for such systems, we could experiment with planets in multi-star systems. This is an exciting prospect since it challenges conventional traditions of planets orbiting a single sun or electrons orbiting a single nucleus.

\section{Final Conclusion} \label{section:conclusion}

This paper presents an investigation of the application of numerical simulation and Bayesian optimisation techniques in the study of celestial dynamics. By utilising a 2D three-body system simulator and Bayesian optimisation methods, we were able to satisfactorily locate two (potentially three) out of five Lagrange points (L4, L5, and perhaps L3) whilst evaluating their stability properties. Additionally, the study delved into the use of techniques to find stable configurations in multi-star systems represented by regular convex \textit{n}-gons. Through search-space simplification and the use of appropriate loss functions, we were able to investigate the dynamics of these systems in greater detail. Overall, the study provides insight into the use of Bayesian optimisation within the celestial dynamics field and highlights the technique's effectiveness.
\newpage 
{
\small
\bibliographystyle{unsrt}
\bibliography{main}
}

\end{document}